\documentclass[12pt,twoside,headsepline,openright]{article}
\usepackage[english]{babel}
\usepackage{graphicx}

\begin{document}
{\noindent \Large \sffamily \bfseries
 Superpositions of the Orbital Angular Momentum\\ for Applications in Quantum Experiments}

\vspace{8mm}

 {\noindent
 Alipasha Vaziri, Gregor Weihs, and Anton Zeilinger}

\vspace{8mm}

 {\noindent \itshape Institut f\"{u}r Experimentalphysik, Universit\"{a}t Wien \\
  \itshape Boltzmanngasse 5, 1090 Wien, Austria}

\vspace{8mm}

\abstract{ Two different experimental techniques for preparation
and analyzing superpositions of the Gaussian and Laguerre-Gassian
modes are presented. This is done exploiting an interferometric
method on the one hand and using computer generated holograms on
the other hand. It is shown that by shifting the hologram with
respect to an incoming Gaussian beam different superpositions of
the Gaussian and the Laguerre-Gaussian beam can be produced. An
analytical expression between the relative phase and the
amplitudes of the modes and the displacement of the hologram is
given. The application of such orbital angular momenta
superpositions in quantum experiments such as quantum cryptography
is discussed.}
\section{Introduction}

In recent years a steadily growing interest in orbital angular
momentum states of light can be observed. These light fields which
are solutions of the scalar wave equation are mathematically
described by Laguerre-Gaussian modes possessing a helical phase
structure. As a result they have one or more phase singularities.
The orbital angular momentum carried by these light fields is
distinct from the angular momentum associated with polarization,
it is quantized in units of $\hbar$ and and can be converted into
mechanical angular momentum \cite{He95a}. The possibility of using
such light fields for driving micromachines, and their application
as optical tweezers and optical traps make them possibly useful
\cite{Simpson97a,Galajda01a,Friese96a,Paterson01a}.

One may be led to believe that the orbital angular momentum of
photon pairs created by spontaneous parametric down-conversion is
not conserved if using classical detection methods\cite{Arlt99a}.
However from the quantum physics perspective most interesting are
those applications which exploit the quantum properties of photons
with orbital angular momentum. As already shown \cite{Mair01a}
spontaneous parametric down-conversion creating pairs of photons
conserves orbital angular momentum on the individual photon level.
Also, the two down-converted photons in a pair have been
demonstrated to be in an entangled state with respect to their
orbital angular momentum. As the Laguerre-Gaussian modes can be
used to define an infinitely dimensional Hilbert space, orbital
angular momentum entangled photons provide access to
multi-dimensional entanglement which involves many orthogonal
quantum states. Multi-dimensional entangled states are of
considerable importance in the field of quantum information and
quantum communication enabling, for example, quantum cryptography
with higher alphabets.

For these quantum applications of orbital angular momentum it is
necessary to be able to analyze orbital angular momentum entangled
states of individual photons and one has to have quantitative
measures for the multi-dimensional entanglement. Since analyzing
entanglement locally always implies analyzing superpositions it is
essential to have experimental techniques for preparing and
analyzing superpositions of different orbital angular momentum
eigenstates. One quantitative measure for multi-dimensional
entanglement is a Bell inequality experiment generalized to more
than two dimensions.

In the following we will present techniques, already realized in
experiment, for preparing superpositions of Laguerre-Gaussian
modes with arbitrary amplitude and phase ratios. This was done
employing two different techniques, using computer generated
holograms on the one hand and an interferometric method on the
other hand. We will also present how these techniques will be
applied in a Bell-inequality experiment as mentioned above, which
is in progress in our laboratory.

\section{Mathematical description of the Laguerre-Gaussian modes}



The well-known Gaussian beam is not the only solution of the
scalar paraxial wave equation. Also, the Hermite Gaussian (HG)
\cite{Saleh91a} and the Laguerre-Gaussian (LG) modes for which the
electromagnetic field amplitude is given by
\begin{equation}
u_{p,l}(r,\theta ,z)= \sqrt{\frac{2p!}{\pi
(p+|l|)!}}\frac{1}{w(z)} \left( \frac{r\sqrt{2}}{w(z)} \right)
^{|l|}L_{p}^{|l|} \left(\frac{2r ^{2}}{w(z) ^{2}}\right)e
^{\frac{-r^{2}}{w(z) ^{2}}}
e^{\frac{-ikr^{2}}{2R(z)}}e^{-i(2p+|l|+1) \arctan(
\frac{z}{z_{R}})}e^{-il \theta }\label{1c}
\end{equation}

are solutions of this equation. With their two-fold infinite
number of the indices both the LG and the HG modes build an
orthogonal basis set for describing any paraxial transversal mode
of the free propagation. For our further considerations we will
only focus on the LG modes.

An LG mode (\ref{1c}) is characterized by its two indices $p$ and
$l$ and by the standard Gaussian beam parameter definitions for
the spot size $w(z)$, the radius of wavefront curvature $R(z)$ and
the Rayleigh length $ z_{R}$. The $L_{p}^{l}(x)$ term in
(\ref{1c}) is a generalized Laguerre polynomial. The indices $p$
and $l$ are referred to as the radial and azimuthal mode index
respectively, $p+1$ is the number of radial nodes and $l2\pi$ is
the phase variation along a closed path around the beam center
(Figure 1). This phase variation which is due to the $e^{-il
\theta }$ term in (\ref{1c}) results in a helical structure of the
wave front. In consequence there is a phase singularity in the
beam center for $l\neq0$  and in order to satisfy the wave
equation the intensity has to vanish there. Therefore such states
are also called doughnut modes. Since the Laguerre-Gaussian modes
(\ref{1c}) are angular momentum eigenstates they carry an orbital
angular momentum of $\hbar l$ per photon. The fact Laguerre
Gaussian modes carry an orbital angular momentum was predicted and
it was experimentally verified \cite{He95a, Friese96a}.

As usual in quantum mechanics, external variables describe the
quantum state in real space while internal variables refer to
additional variables. In that sense it is important to stress that
the angular momentum carried by Laguerre Gaussian modes with
$l\neq 0$ is an external angular momentum distinct from the
internal angular momentum of the photons associated with their
spin \cite{Allen92a}.

\section{Production of the Laguerre-Gaussian modes}

There are several experimental methods like cavity induced
production, astigmatic mode conversion \cite{Beijersbergen93a} and
the use of computer generated holograms \cite{Arlt98a} for
creating LG modes. In this article we restrict ourselves to
describing the use of computer generated holograms.

A hologram is a recording of the interference pattern between the
desired field and some reference field. The simplest possible
reference field is the plane wave.
\begin{equation}
R = R_{0}e^{ik_{x}x+ik_{z}z} \label{1d}
\end{equation}
The interference pattern produced by such a beam propagating at an
angle $\xi=\arctan( \frac{k_{x}}{k_{z}})$ and e.g. an $LG_{01}$
mode propagating along the z-direction can be calculated
numerically (Figure 2a). It is a line grating with one dislocation
in the center. Now, if this grating is illuminated by the
reference beam, which is sufficiently well realized if this
hologram is placed at the waist of a Gaussian beam, the $LG_{01}$
mode is reproduced. Intuitively speaking the phase dislocation
exerts a ``torque'' onto the diffracted beam because of the
difference of the local grating vectors in the upper and lower
parts of the grating. This ``torque'' depends on the diffraction
order n and on the number of dislocations $\Delta m$ of the
hologram. Consequently the right and left diffraction orders gain
different handedness and the associated orbital angular momentum
values differ in their sign. The definition of the sign can be
chosen by convention. The modulation of the incoming beam can be
either done in the phase using transmission phase gratings or
reflection gratings or in the amplitude using absorption gratings
the latter being rather inefficient.

In our experiments we used transmission phase gratings with a
diffraction efficiency of 70$\%$ after blazing. The binary
structure of the hologram in (Figure 2a) is modified by blazing
such that it results in a pattern as shown in (Figure 2b). The
transmission function of such a hologram in polar coordinates is
\cite{Arlt99b}

\begin{equation}
T(r,\phi) = e^{i \delta \frac{1}{2\pi}\mathrm{mod}(\Delta m\phi -
\frac{2\pi}{\Lambda}r \cos \phi, 2 \pi)} \label{1e}
\end{equation}

where $\delta$ is the amplitude of the phase modulation and the
second factor in the superscript is the actual pattern of the
blazed hologram. $\Lambda$ is the spacing period of the grating
and $\mathrm{mod}\;(a,b)=a-b\;\mathrm{Int} \left ( \frac{a}{b}
\right )$. As mentioned above the fraction of intensity diffracted
into higher orders of the hologram that is for a grating with one
dislocation to the higher order LG modes with $l\neq1$ can be
decreased by blazing. But the blazing has no influence on the
composition of the output beam's radial index p-terms. The
relative amplitudes of the p-terms depend on the choice of the
beam parameters of the input beam to output beam \cite{Arlt99b}.
However for our experimental applications using holograms with one
dislocation the relative amplitudes of the $p\neq 0$-terms become
negligible.

\section{Superpositions of the Gaussian and the Laguerre-Gaussian
modes}

Central to many, if not all, quantum experiments is the concept of
superposition. It is therefore important to be able to both
produce and analyze superpositions of the various states of a
chosen basis. We therefore discuss now superpositions of the LG
modes presented above. There are several experimental methods for
producing such superpositions. One simple way is to use a
Mach-Zehnder interferometer as sketched in (Figure 3a). After the
input mode is split by the first beam splitter the beam of each
arm is sent through a hologram causing the desired mode
transformations which are in general different transformations in
the two arms. For producing superpositions of the Gaussian and the
$LG_{01}$ mode it is sufficient to place a hologram with one
dislocation only in one of the arms of the interferometer. The two
beams are brought together on a second beam splitter where they
are superposed. An advantage of this method is that by attenuating
each of the arms and using a phase plate superpositions with
arbitrary amplitudes and relative phases can be produced without
changing the setup. The resulting interference pattern for a
superposition of an $LG_{00}$ mode (=Gaussian mode) and an
$LG_{01}$ mode is shown in Figure 3b.

However the interferometric preparation of superposition modes has
also some disadvantages. The experimental setup becomes too
complex and too difficult to control when one has to create and
analyze superpositions several times. It would be necessary to
keep the relative phase in the interferometer arms stable and one
also has to take the different divergences of the interfering
modes into account. A more convenient but less general method for
creating superposition modes is to use a displaced hologram
\cite{Mair01a}. This method is particularly suitable for producing
superpositions of an $LG_{0l}$ with the Gaussian mode which may
also be seen as an $LG$ mode with $l=0$. The transmission function
of a hologram which is designed to transform a Gaussian mode into
an $LG_{01}$ mode is given by (\ref{1e}) with $\Delta m=1$. In
order to tranform an incoming Gaussian beam into an $LG_{01}$ beam
the hologram should be placed at the waist of the Gaussian beam
and the beam should be sent through the center of the hologram
where the dislocation is located. The intensity pattern of such an
$LG_{01}$ mode posseses a centrally located singularity. By
shifting the dislocation out of the beam center step by step one
can experimentally observe that the singularity becomes eccentric
resulting in the same pattern achieved by the interferometric
setup.

Numerical simulations show that there are also higher order
$LG_{0l}$ components present in the case when the superposition is
produced by a displaced hologram. The traces in the bottom of
Figure 5c are corresponding to the amplitudes of the $LG_{0-1}$,
$LG_{02}$ and $LG_{0-2}$ modes. This is also required by the
unitary of the procedure. Nevertheless the relation (\ref{1g})
between the relative amplitudes and the position of the
singularity is still a good approximation because the amplitudes
of these higher orders are small. In the case where the beam is
sent through a border region of the hologram far away from the
dislocation the output beam is again a Gaussian beam because there
the hologram acts as an ordinary grating.

An important question is whether in the experiment one actually
observe coherent superpositions rather than incoherent mixtures.
The distinction between coherent superposition and incoherent
mixture of Gaussian and LG modes is that the latter posses no
phase singularity. This is because adding the spatial intensity
distributions of these two modes will yield a finite intensity
everywhere in the resulting pattern. In contrast, in a coherent
superposition the amplitudes are added and therefore the phase
singularity must remain and is displaced to an eccentric location
(Figure 3b). It will appear at that point where the amplitudes of
the two modes are equal with opposite signs. Therefore the radial
distance of the singularity from the beam center is a measure of
the amplitude ratio of the Gaussian to the $LG$ components whereas
the angular position of the singularity is determined by their
relative phase.

To obtain quantitative results we calculated the intensity
distribution of a normalized superposition mode of an $LG_{00}$
and an $LG_{01}$ mode described by
\begin{equation}
 \frac{1}{\sqrt{1+\gamma^{2}}}\left[|u_{00}\rangle+\gamma e^{i\varphi}|u_{01}\rangle\right]\label{1f}
\end{equation}
and looked for the position of its phase singularity. Here
$|u_{00}\rangle$ and $|u_{01}\rangle$ denote the amplitudes of an
$LG_{00}$ and an $LG_{01}$ respectively as given by (\ref{1c}),
$\gamma$ is the relative amplitude of the $LG_{01}$ mode and
$\varphi$ is the relative phase of the interfering modes. After
inserting the corresponding $LG$-amplitudes (\ref{1c}) into
(\ref{1f}) we found for the position of the singularity the
cylindrical coordinates
\begin{equation}
r=\frac{w_{0}}{\gamma\sqrt{2}}\mbox{,}\;\; \theta=\varphi,
\label{1g}
\end{equation}
 where $w_{0}$ denotes the waist size of the Gaussian
beam.(Figure 3b).

Although this expression only holds for superpositions of
$LG_{00}$ and $LG_{0\pm1}$ modes it can easily be generalized to
superpositions of higher order LG mode containing more than just
one phase singularity.

In order to prove that the eccentric mode is indeed a
superposition mode one has to project the superposition onto the
orthogonal basis states. Experimentally this was done by sending a
Gaussian laser beam (HeNe, 632nm) through a displaced hologram
producing a superposition of the Gaussian and the $LG_{01}$ mode.
In the next step the output mode was projected onto the Gaussian
mode by coupling into a mono-mode optical fiber. Since all other
modes have a larger spatial extension, only the Gaussian mode can
propagate in the mono-mode fiber and it therefore acts as a filter
for the $LG_{0l}$ modes with $l\neq0$.

Having only a filter transmitting the Gaussian mode the $LG_{01}$
mode had to be identified via an additional step. This was done by
introducing a second hologram making a mode transformation on the
output beam reducing the azimuthal index $l$ by one before
coupling into mono-mode optical fibers(Figure 4). For each
position of the displaced hologram the transmitted intensity to
the Gaussian and the $LG_{01}$ detector was measured.

The results are shown in Figure 5a and 5b. When the hologram is
centered the incoming Gaussian mode is transformed into an
$LG_{01}$ mode. As a result the intensity of light coupled into
the Gauss-detector is a minimum and the one coupled into the
$LG_{01}$-detector a maximum. As the hologram is shifted the
intensity at Gauss-detector increases and the intensity at the
$LG_{01}$-detector drops. The asymmetry in Figure 5a) is a result
of the imperfection of the holograms \cite{jena}. However the
extinction ratio $e$ for all measurements was always far better
than $1:300$.

These results are in agreement with our numerical calculations of
the superposition modes(Figure 5c). The action of the hologram on
the incoming field is characterized by (\ref{1e}). The transmitted
beam directly after the hologram is given by
\begin{equation}
u_{out}(r,\phi) = T(r,\phi)u_{in}(r,\phi)\label{1h}
\end{equation}
Denoting the relative position of the hologram with respect to the
beam center with $(r_{0}, \phi_{0})$ one finds the projection of a
transmitted $LG_{0l}$ mode onto the $LG_{0L}$-mode to be

\begin{equation}
a_{L}^{l}(r_{0},\phi_{0}) = \int_{-\infty}^{+\infty}
\int_{0}^{2\pi} r d r d\phi \left(u_{0L}(r, \phi,0)e ^{-i\Delta
m\frac{2\pi}{\Lambda}r\cos \phi}\right) ^{*}
T(r-r_{0},\phi-\phi_{0}) u_{l}(r,\phi,0) \label{1i}
\end{equation}

This expression represents the amplitude of an $LG_{0L}$-mode when
a hologram with one dislocation placed at $(r_{0}, \phi_{0})$ is
illuminated by an $LG_{0l}$ mode. The numerical simulation of a
setup which corresponds to our experiment is given in Figure 5c.

As mentioned above, the superposition does not only consist of two
modes. The output mode will also contain higher order modes
(Figure 5c bottom). This is a necessary consequence of the fact
that the action of the hologram is unitary. In general, the
relative amplitudes of the higher order modes will increase when
the hologram has many dislocations or when the single dislocation
hologram is illuminated by an $LG_{0l}$ beam with $|l|>1$. However
they are negligible (Figure 5c bottom) for the case that the
hologram is illuminated by an $LG_{01}$ mode.

\section{Conclusion and Outlook}

In this work we showed techniques for producing and analyzing
superpositions of Laguerre-Gaussian modes. Our experimental
results show that it is possible to achieve mode detection of high
distinction ratio ($1:300$) with the technique described above. As
recently shown the orbital angular momentum of photons is
conserved in parametric down-conversion \cite{Mair01a} and the
down-converted photons are found to be in an entangled state with
respect to the orbital angular momentum. Since orbital angular
momentum entangled photons can be used as qu-nits they open a
practical approach to multi-dimensional entanglement where the
entangled states do not only consist of two orthogonal states but
of many of them. We expect such states to be of importance for the
current efforts in the field of quantum computation and quantum
communication. For example, quantum cryptography with higher
alphabets could enable one to increase the information flux
through the communication channels.

However the ultimate confirmation of the entanglement of the
orbital angular momentum will be a Bell inequality experiment
generalized to more states \cite{Kaszlikowski00a}. Such an
experiment will also give a quantitative measure of the
multi-dimensional entanglement produced by parametric
down-conversion. Employing the techniques presented here for
preparing and analyzing superpositions of orbital angular momentum
states such an experiment is presently in progress in our
laboratory.

{\bf\large Acknowledgment}

The Authors would like to thank Marek Zukowski for discussion and
Jochen Arlt for correspondence. This work was supported by the
Austrian Fonds zur F\"{o}rderung der wissenschaftlichen Forschung
(FWF).

\clearpage

\clearpage
\begin{figure}
\begin{center}
\includegraphics[height=0.7\textwidth,angle=0]{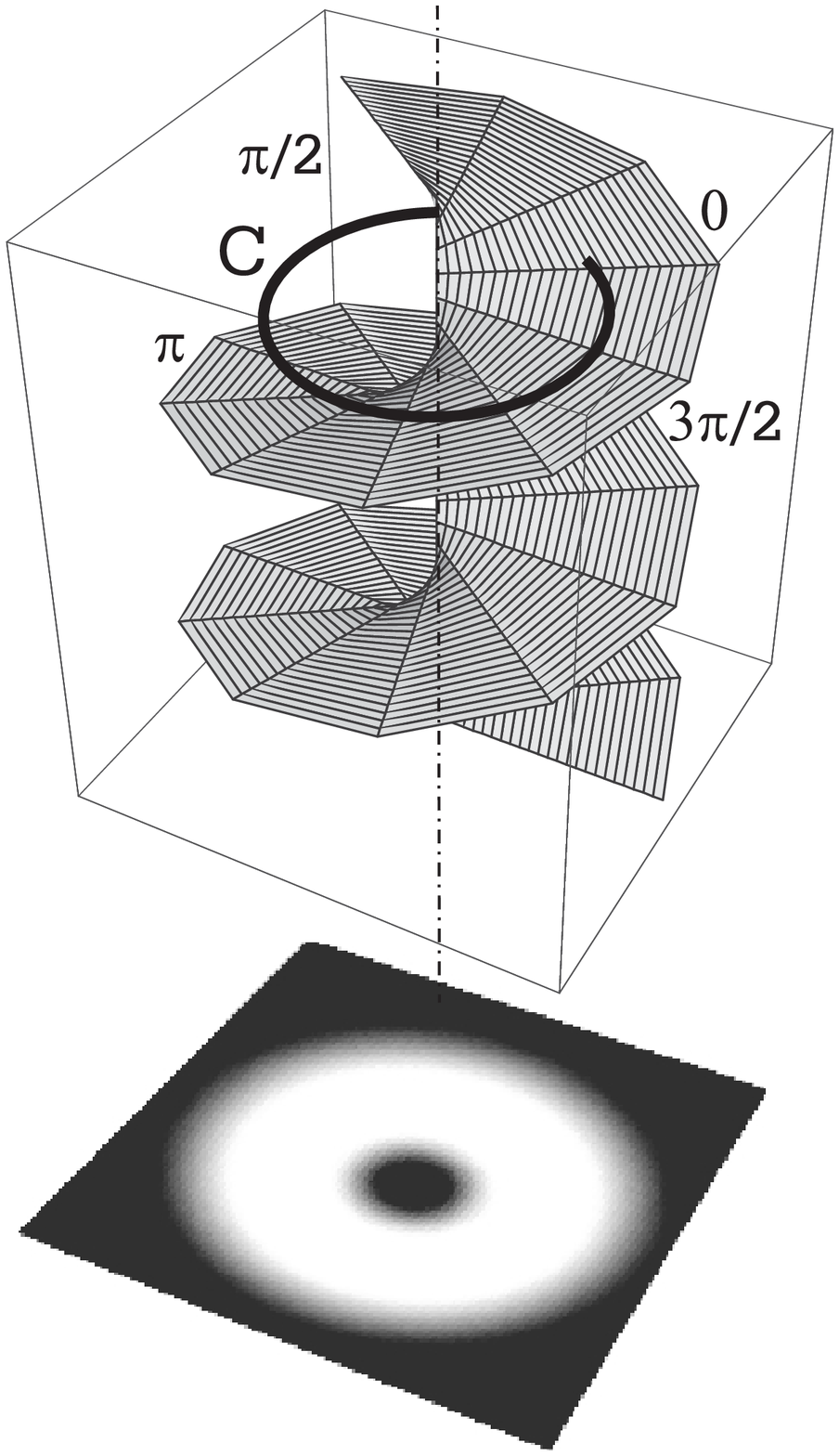}
\end{center}
\caption{The wave front (top) and the intensity pattern (bottom)
of the simplest Laguerre-Gaussian (LG) or doughnut mode. The
azimuthal phase term $ e^{-il \theta}$ of the LG modes results in
helical wave fronts. The phase variation along a closed path is $
2 \pi l$. Therefore in order to fulfill the wave equation the
intensity has to vanish in the center of the beam.}
\label{vaziri_fig1}
\end{figure}

\begin{figure}
\begin{center}
\includegraphics[height=0.5\textwidth,angle=0]{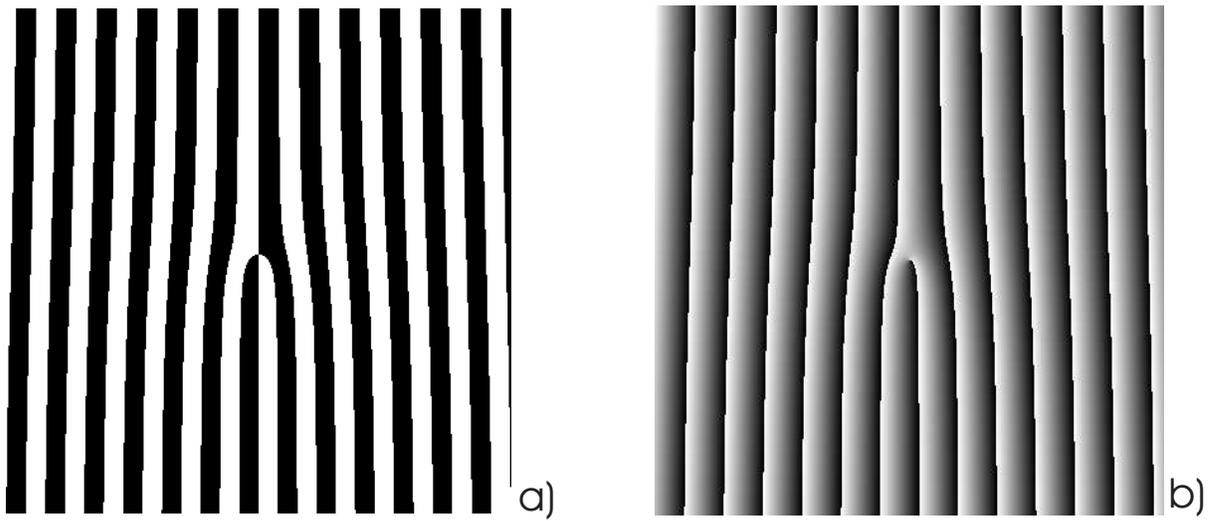}
\end{center}
\caption{Computer generated binary a) and blazed b) templates for
computer generated holograms with single dislocation. By
illuminating these gratings with a Gaussian beam an $LG_{01}$ mode
is produced in their first diffraction order. The diffraction
efficency in the desired order can be increased by blazing.}
\label{vaziri_fig2}
\end{figure}

\begin{figure}
\begin{center}
\includegraphics[width=\textwidth,angle=0]{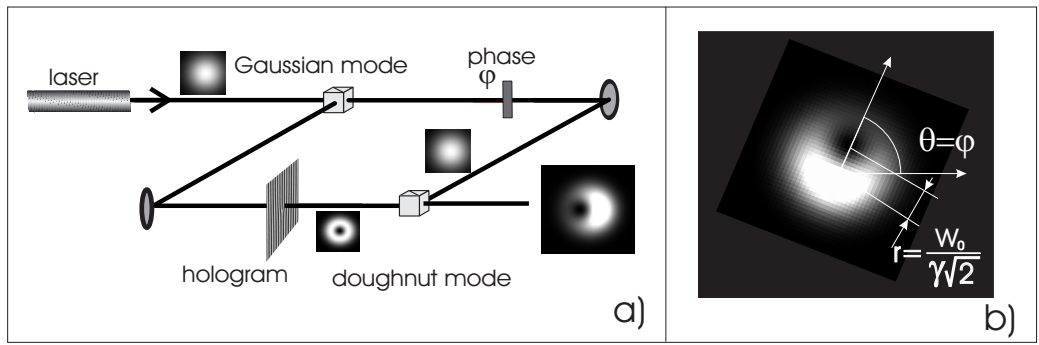}
\end{center}
\caption{ Superpositions of the Gaussian and the $LG_{01}$ mode.
Using an interferometer a) with a single dislocation hologram
placed in one arm superpositions of the Gaussian and the $LG_{01}$
can be produced. Such superpositions posses an eccentric
singularity b) where the radial distance of of the singularity
from the beam center is a measure of the amplitude ratio of the
Gaussian to the LG components and the angular position of the
singularity is determined by their relative phase $ \varphi$.}
\label{vaziri_fig3}
\end{figure}

\begin{figure}
\begin{center}
\includegraphics[width=\textwidth,angle=0]{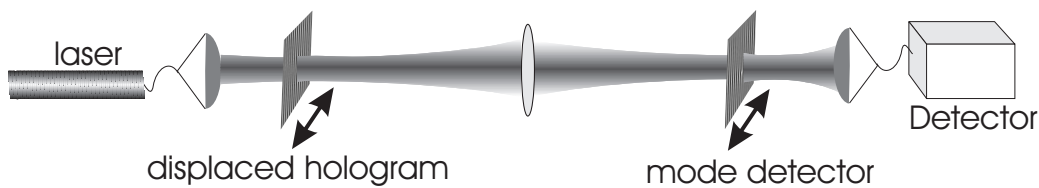}
\end{center}
\caption{Measurement of the Gauss and the $LG_{01}$ components of
a superposition mode. The displaced hologram produces a
superposition of the Gaussian and the $LG_{01}$ mode. The relative
amplitudes are determined by The mode detector which consists of a
second hologram and a mono-mode optical fiber makes a scan
determining the relative amplitudes of the Gaussian and the
$LG_{01}$ mode.} \label{vaziri_fig4}
\end{figure}

\begin{figure}
\begin{center}
\includegraphics[width=0.5\textwidth,angle=0]{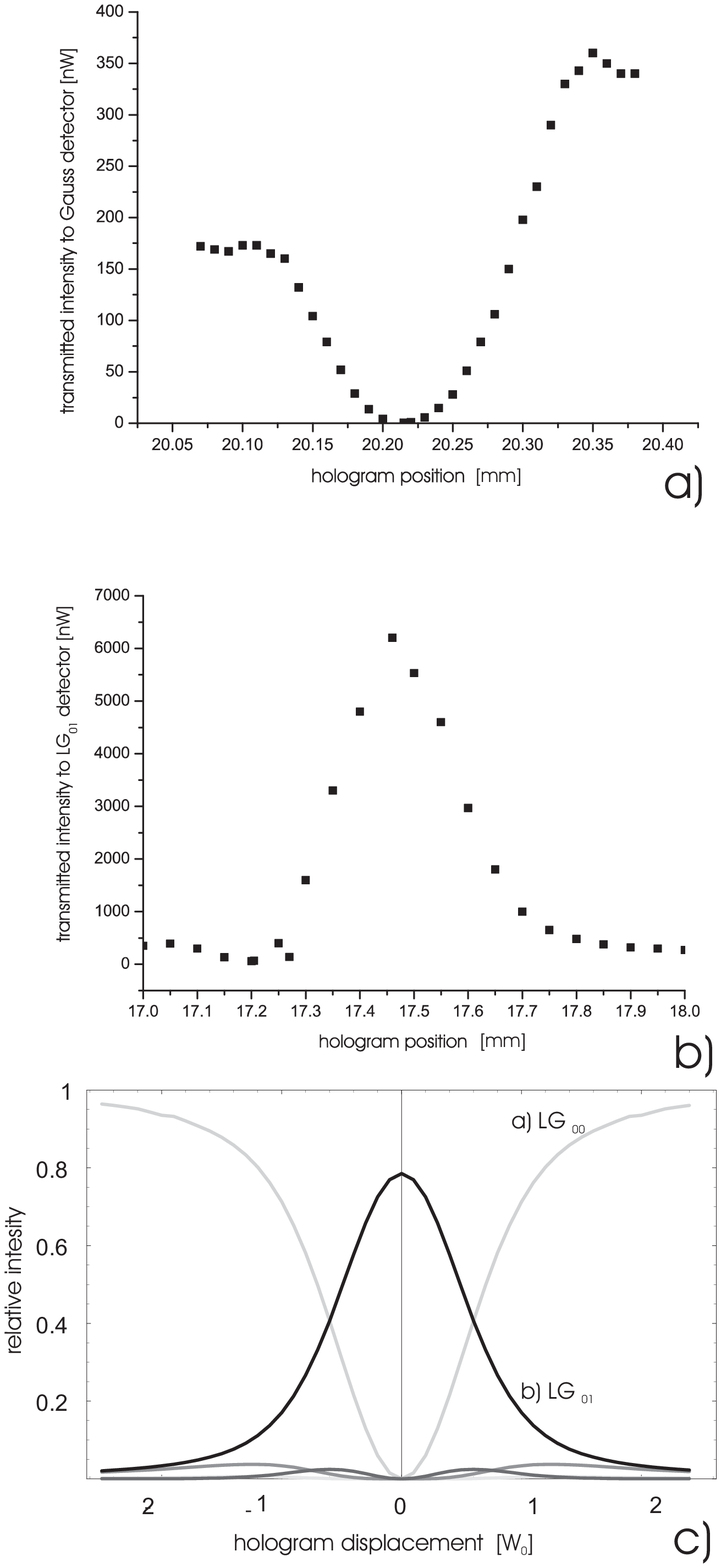}
\end{center}
\caption{Mode decomposition after a displaced hologram;
experimental results 5a), 5b) and simulation 5c). Superpositions
of Gaussian and $LG_{01}$ modes were produced by a displaced
hologram. For each displacement the intensities at the Gauss-5a)
and at the $LG_{01}$-5b) detector were measured. The same results
were also achieved in numerical simulation 5c) where a) matches
5a) and b) 5b). The traces in the bottom of 5c) correspond to the
amplitudes of the $LG_{0-1}$, $LG_{02}$ and $LG_{0-2}$ modes which
are negligible in this experiment. The unsymmetry in 5a) is due to
the imperfection of the hologram. } \label{vaziri_fig5}
\end{figure}

\clearpage
\newpage
\bibliographystyle{unsrt}

\end{document}